\documentclass[conference,a4paper]{IEEEtran}
\IEEEoverridecommandlockouts
\usepackage{times}
\usepackage{amsmath,amssymb,bm,amsfonts,enumerate,url,footnote,color,ifpdf}
\usepackage[noend]{algpseudocode}
\ifCLASSINFOpdf
	\usepackage[pdftex]{graphicx}
\else
	\usepackage[dvips]{graphicx}
\fi
\usepackage{array,multirow,cite}
\ifCLASSOPTIONcompsoc
	\usepackage[caption=false,font=footnotesize,labelfont=sf,textfont=sf]{subfig}
\else
	\usepackage[caption=false,font=footnotesize]{subfig}
\fi
\usepackage[english]{babel}
\usepackage{booktabs,rotating,balance,arydshln} 
\usepackage{tikz} \usetikzlibrary{matrix}
\usepackage{algorithm}
\usepackage[left=1.36cm,right=1.36cm,top=1.83cm, bottom=4.45cm]{geometry}

\usepackage{booktabs} 
\usepackage{tabularx} 
\usepackage{setspace}
\usepackage{titlesec}

\usepackage{titlesec}
\titlespacing\section{0pt}{10pt plus 4pt minus 2pt}{4pt plus 2pt minus 2pt}
\titlespacing\subsection{0pt}{8pt plus 4pt minus 2pt}{2pt plus 2pt minus 2pt}

\begin{document}

\title{Digital Twin-Empowered Deep Reinforcement Learning for Intelligent VNF Migration in Edge-Core Networks}
\author{
    \IEEEauthorblockN{Faisal Ahmed\IEEEauthorrefmark{1}, Suresh Subramaniam\IEEEauthorrefmark{3}, Motoharu Matsuura\IEEEauthorrefmark{7}, Hiroshi Hasegawa\IEEEauthorrefmark{8}, \\ and Shih-Chun Lin\IEEEauthorrefmark{1}}
    \IEEEauthorblockA{\IEEEauthorrefmark{1}North Carolina State University, Raleigh, NC, USA,
     \{fahmed5, slin23\}@ncsu.edu}
    \IEEEauthorblockA{\IEEEauthorrefmark{3}The George Washington University, Washington, USA, 
    suresh@gwu.edu}
    {
}
    \IEEEauthorblockA{\IEEEauthorrefmark{7}University of Electro-Communications, Chofu, Japan, 
    m.matsuura@uec.ac.jp}
    \IEEEauthorblockA{\IEEEauthorrefmark{8}Nagoya University, Nagoya, Japan, 
    hasegawa@nuee.nagoya-u.ac.jp}
    
    \thanks{This work was supported in part by the National Science Foundation (NSF) under Grant CNS-221034, Meta 2022 AI4AI Research, and the NC Space Grant.}
}
\setlength{\baselineskip}{10.05pt} 

\maketitle
\vspace{-10pt}
\begin{abstract}
The growing demand for services and the rapid deployment of virtualized network functions (VNFs) pose significant challenges for achieving low-latency and energy-efficient orchestration in modern edge–core network infrastructures. To address these challenges, this study proposes a Digital Twin (DT)-empowered Deep Reinforcement Learning framework for intelligent VNF migration that jointly minimizes average end-to-end (E2E) delay and energy consumption. The VNF migration task is formulated as a Markov Decision Process and optimized using an Advantage Actor–Critic architecture, enabling adaptive migration decisions under dynamic network conditions. A key innovation of the proposed framework is the integration of a DT module composed of a multi-task variational autoencoder serving as the generative model and a multi-task long short-term memory network serving as the predictive model. Together, these models simulate environment dynamics and generate high-quality synthetic experiences that enhance training efficiency and accelerate policy convergence. Simulation results demonstrate substantial improvements in both end-to-end delay and energy consumption, establishing a new benchmark for intelligent VNF migration in edge–core networks.
\end{abstract}
\begin{IEEEkeywords}
Digital Twin, Advantage Actor-Critic, VNF Migration, Edge-Core Network, VNF-FG
\end{IEEEkeywords}
\IEEEpeerreviewmaketitle

\section{Introduction}

Network Function Virtualization (NFV) \cite{yi2018comprehensive} marks a paradigm shift in network architecture, designed to lower operational costs by decoupling network functions (NFs) from specialized hardware appliances, such as conventional middle-boxes. By abstracting these NFs into software-based entities, namely, Virtualized Network Functions (VNFs) \cite{8556457}, NFV enables their deployment on general-purpose commodity servers. These deployments typically consist of ordered chains of VNFs including elements such as load balancers and firewalls.
This architectural flexibility facilitates faster service provisioning, enhances scalability, optimizes resource utilization, and reduces operational expenditures for service providers \cite{liang2022low}.

However, as the adoption of NFV expands and the number of VNFs increases, scaling the VNF forwarding graph (VNF-FG) within edge networks becomes increasingly challenging due to their limited computational resources. In contrast, core networks typically possess abundant computing capacity \cite{9440734}, making them well-suited for offloading VNFs from the edge. Integrating edge and core infrastructures into a unified edge–core network architecture thus offers enhanced flexibility for optimizing end-to-end delay (E2E) and improving overall service performance \cite{9440734}. 
Nevertheless, this shift toward large-scale, distributed architectures brings forth additional challenges, most notably, the need to ensure a stable and reliable power supply. Given that energy consumption accounts for a significant share of operational expenditure, its intelligent and adaptive management is essential for achieving sustainable network growth \cite{9472805}. In response, Network Service Providers are actively pursuing strategies to reduce the energy consumption of these infrastructures, with the overarching goal of enhancing operational efficiency \cite{10575040}.

Recently, significant research efforts have focused on optimizing the scheduling of VNFs, particularly deployment and migration strategies \cite{9440734, 8932445, liang2022low} to minimize E2E delay \cite{liang2022low} and reduce energy consumption \cite{10509644,9472805}. Various studies have proposed deployment strategies aimed at optimizing delay and energy consumption \cite{liang2022low, 10575040}; however, these often rely on static VNF placement, assuming fixed network demands and unconstrained resource availability. Such assumptions undermine their applicability in real-world scenarios, where demand is dynamic and resource contention is common. To overcome these limitations, VNF migration \cite{liang2022low, 8905349} has emerged as a more adaptive and robust solution. By enabling the flexible relocation of VNFs to more optimal servers while preserving application states, VNF migration allows network infrastructures to respond dynamically to changing workloads and resource conditions, ensuring service continuity and efficient resource utilization in heterogeneous environments \cite{liang2022low}.

In parallel, the increasing complexity and dynamism of modern networks have fueled interest in Machine Learning (ML) for intelligent network management \cite{xie2018survey}. Although widely used, supervised ML often requires labeled data and struggles with uncertainty. In contrast, many networking tasks involve sequential decision-making under uncertainty, making Reinforcement Learning (RL) a more suitable paradigm \cite{8932445}. Specifically, RL has proven effective in optimizing the orchestration of VNF-FGs across heterogeneous infrastructures, particularly in response to fluctuating service demands. Building upon this foundation, Deep Reinforcement Learning (DRL)\cite{8932445} combines RL with deep neural networks (DNNs) to tackle complex problems in high-dimensional spaces. DRL has shown great potential in tasks such as VNF migration in edge and core networks, where real-time decisions and efficient use of resources are crucial. 

Although DRL has demonstrated notable performance improvements, it is typically implemented to solve large action spaces in tasks such as VNF migration and requires a substantial amount of training data. This, in turn, necessitates frequent interactions with the physical environment to collect sufficient data, imposing considerable overhead on real-world systems. To mitigate the burden a Digital Twin (DT) can be implemented to simulate environment dynamics and generate additional training data \cite{10522623}. By serving as a virtual replica of the physical system, the DT enables the creation of a virtual experience buffer that, when combined with the physical experience buffer enriches the training dataset and accelerates policy convergence while reducing real-world exploration costs. Furthermore, hosting DRL training processes within the DT can offload computational workloads from the physical system, thereby reducing energy consumption and minimizing disruptions caused by continuous model training.

Building on the aforementioned insights, this study proposes a novel DT-empowered DRL framework for VNF migration in edge–core networks.
The key contributions of this work are summarized as follows:

\begin{itemize}

\item To minimize the E2E delay of VNF-FGs and reduce the overall energy consumption of edge–core networks, this work proposes a DRL framework based on an Advantage Actor–Critic (A2C) architecture that intelligently migrates VNFs within the edge–core infrastructure. 

\item To enhance the performance and sustainability of the DRL framework for intelligent VNF migration, a novel integration with a DT is proposed that generates synthetic experience data and hosts the A2C training process. By shifting training from the physical environment to the DT, the framework significantly reduces reliance on real-world interactions and minimizes the energy cost and operational disruptions associated with continuous model updates.

\item The DT module incorporates a multi-task Variational Autoencoder (VAE) as the generative model and a multi-task Long Short-Term Memory (LSTM) network as the predictive model. The generative model first produces simulated state–action pairs, and the predictive model then uses these to generate the corresponding simulated next states and reward values. Together, these components enrich the training dataset with synthetic transitions stored in the DT experience buffer, thereby improving the training efficiency of the A2C model.


\item Extensive simulations demonstrate that the proposed framework significantly outperforms baseline methods. For example, in terms of average E2E delay and average energy consumption, the proposed framework achieves reductions of 7\% and 15\%, respectively, compared to baselines such as non DT-based A2C VNF migration.
\end{itemize}


\label{sec:introduction}

\section{System Model and Problem Formulation}
\begin{figure} [t]
\centerline{\includegraphics[width=0.95\columnwidth]{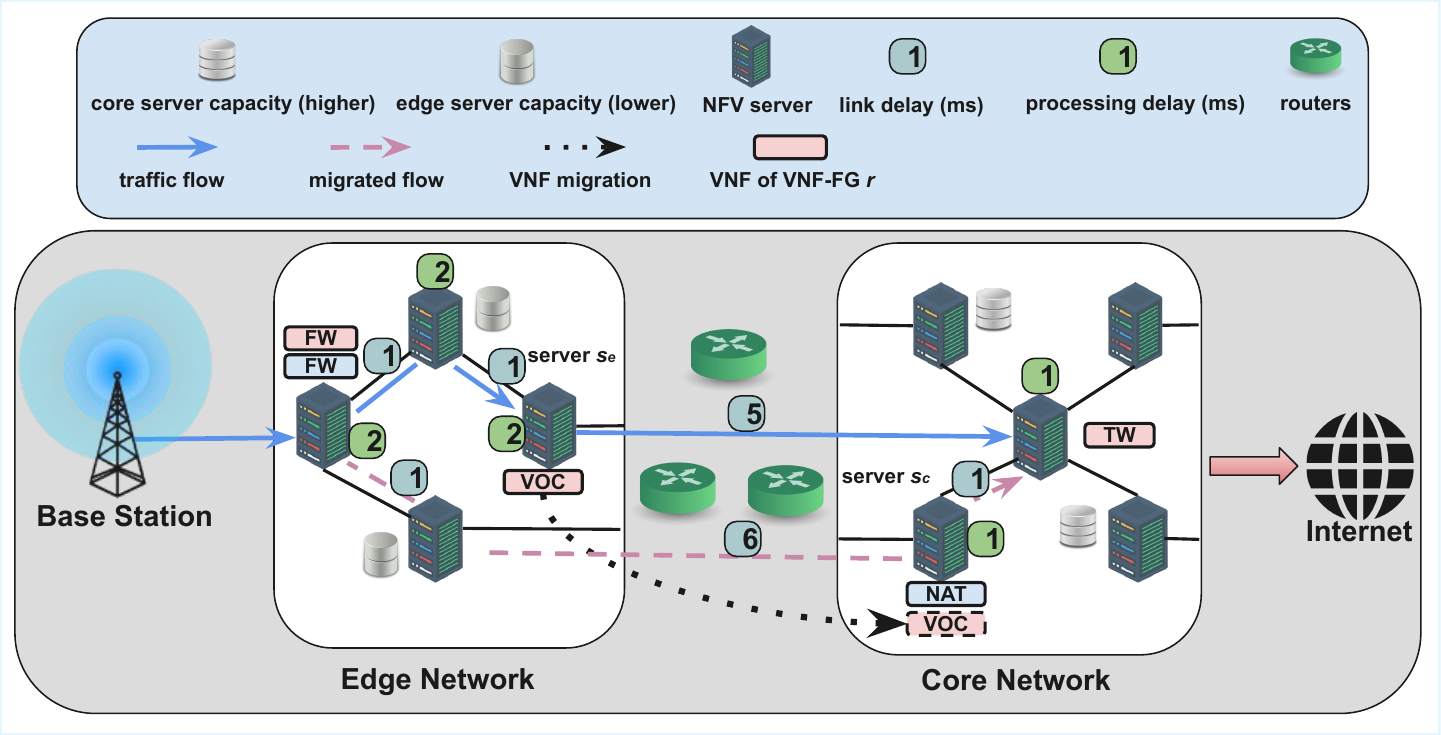}}
   \caption{A representative example of an 8K video streaming service served by a VNF-FG demonstrates that the E2E delay is reduced from 14 ms to 12 ms, and the edge server $s_{e}$ is powered off following VNF migration, thereby reducing overall energy consumption.}
\end{figure}
The considered edge–core network is modeled as a weighted undirected graph $\Omega = (\mu, \nu)$, where $\mu$ denotes the set of servers and $\nu$ represents the set of connectivity links between them. The server set is partitioned as $\mu = \mu_e \cup \mu_c$, where $\mu_e = \{s_{1}, s_{2}, \dots, s_{e}\}$ represents the subset of edge cloud servers, and $\mu_c = \{s_{e+1}, \dots, s_{c}\}$ corresponds to the subset of core cloud servers. Here, $s_1 \leq s \leq s_e$ indexes the edge cloud servers, and $s_{e+1} \leq s \leq s_c$ indexes the core cloud servers.
The network architecture is assumed to include multiple levels of switching infrastructure to facilitate inter-server connectivity. Each link $l \in \nu$ connects a pair of distinct servers, such as $s_{e+1}$ and $s_{c}$, where $s_{e+1}, s_{c} \in \mu_{c}$ and $s_{e+1} \ne s_{c}$. Each server $s \in \mu$ is characterized by its available computational resources, namely memory and CPU. Let $\zeta_{s}^{{MEM}}$ and $\zeta_{s}^{{CPU}}$ denote the memory and CPU capacities of server $s$, respectively. It is assumed that core cloud servers offer greater computational resources, whereas edge servers operate with comparatively limited capacity. Additionally, each connectivity link $l = (s_{e+1}, s_{c}) \in \nu$ has an associated available bandwidth denoted by $\zeta_{l}^{{BW}}$. 

To accommodate dynamic network conditions, the NFV system is assumed to operate in a time-slotted manner. Each service request, represented as a VNF-FG $r \in R$, comprises a set of VNFs, denoted by $V_r = \{v_1, v_2, \dots, v, \dots, v_p \}$, with each VNF deployed on each $s \in \mu$ according to the strategy defined in \cite{8905349}. At each time step $t$, each VNF $v$ has an associated resource demand characterized by memory, CPU, and bandwidth requirements. The memory and CPU demands of VNF $v$ in request $r$ are denoted by $\chi_{r, v}^{\text{MEM}}(t)$ and $\chi_{r, v}^{\text{CPU}}(t)$, respectively. In addition, the bandwidth requirement between any pair of connected VNFs, such as $v_1$ and $v_2$, is represented by $\chi_{r, \overline{l}}^{\text{BW}}(t)$, where $\overline{l}$ denote a logical link between the two VNFs and $\overline{l} \in \bar{L_{r}}$.

\subsection{VNF and Logical Link Mapping in VNF-FG, Associated Constraints, and Service Status}

To begin with, two binary variables are introduced to represent the mapping status of VNFs and logical links within a VNF-FG $r$ at time step $t$.

\begin{equation}
y_{{r},v}^{s}(t) = 
\begin{cases}
1, & \text{server $s$ serves VNF ${v}$ at $t$} \\
0, & \text{otherwise}.
\end{cases}
\end{equation}

\begin{equation}
z_{r, \overline{l}}^{l}(t) = 
\begin{cases}
1, & \text{link $l$ serves logical link $\bar{l}$ at $t$} \\
0, & \text{otherwise}.
\end{cases}
\end{equation}

Even if a server $s$ hosts at least one VNF, it must be activated to perform its functions; the same applies to a link $l$. Therefore, binary variables are introduced to indicate whether a server $s$ and a link $l$ are active at time step $t$.
\begin{equation}
x^{s}(t) = 
\begin{cases}
1, & \sum_{r \in R} \sum_{{v} \in V_r} y_{{r},v}^{s}(t) \geq 1, \; s \in \mu \\
0, & \text{otherwise}.
\end{cases}
\end{equation}

\begin{equation}
x^{l}(t) = 
\begin{cases}
1, & \sum_{r \in R} \sum_{\bar{l} \in \bar{L_r}} y_{{r},v}^{s}(t) \geq 1, \; l \in \nu \\
0, & \text{otherwise}.
\end{cases}
\end{equation}

To ensure that sufficient resources are available on both servers and links when servicing VNF-FGs, the following constraints are defined:
\begin{equation}
\sum_{r \in R} \sum_{{v} \in V_r} y_{{r},v}^{s}(t) \cdot \chi_{r, v}^{{CPU}}(t) \leq \zeta_{s}^{{CPU}}, \quad s \in \mu
\end{equation}
\begin{equation}
\sum_{r \in R} \sum_{{v} \in V_r} y_{{r},v}^{s}(t) \cdot \chi_{r, v}^{{MEM}}(t) \leq \zeta_{s}^{{MEM}}, \quad s \in \mu
\end{equation}
\begin{equation}
\sum_{r \in R} \sum_{ \bar{l} \in \bar{L_r}} z_{r, \overline{l}}^{l}(t) \cdot \chi_{r, \overline{l}}^{{BW}}(t) \leq \zeta_{l}^{{BW}}, \quad l \in \nu
\end{equation}

Each VNF-FG $r \in R$ is associated with a service time $\sigma_{r}^{\text{serv}}$, representing the duration for which $r$ is expected to remain active, provided that all its constituent VNFs are successfully placed while satisfying all system constraints, otherwise, $\sigma_{r}^{{serv}}$ is set to zero. Accordingly, the service status of VNF-FG $r$ at time step $t$ is defined as:
\begin{equation}
\Phi_{r} (t) = 
\begin{cases}
1, & \sigma_r^{ariv} \leq t < \left(\sigma_r^{ariv} + \sigma_r^{serv}\right), \forall t \in T, r \in R \\
0, & \text{otherwise},
\end{cases}
\end{equation}
where, $\sigma_r^{ariv}$ denotes the arrival time of the VNF-FG $r$.

\subsection{E2E Delay Model}
The E2E delay experienced by a VNF-FG consists of three main components: propagation delay, transmission delay, and processing delay. Drawing on the studies in \cite{10729270} and \cite{8532318} the delay incurred on a physical link $l$ for a VNF-FG $r$ at time step $t$ can be expressed as:
\begin{equation}
D^{l}_{link}(t) = D^{l}_{trans}(t) + D^{l}_{prop}(t), \quad l \in \nu, \; r \in R
\end{equation}
here, $D^{l}_{{trans}}(t)$ and $D^{l}_{{prop}}(t)$ represent the transmission delay and propagation delay, respectively, at time step $t$.

Moreover, the processing delay at server $s$ at time step $t$ can be denote as:
\begin{equation}
D^{s}_{proc}(t) = \frac{\Gamma^s_{CPU}(t) \cdot \tau}{1 - \Gamma^s_{CPU}(t)}, \quad s \in \mu, \; r \in R 
\end{equation}
where, $\tau$ denotes the baseline processing delay per flow, and $\Gamma^s_{\text{CPU}}(t)$ is the CPU utilization of server $s$ at time step $t$, defined as:
\begin{equation}
\Gamma^s_{CPU}(t) = \frac{\sum\limits_{r \in R} \sum\limits_{{v} \in V_r} y_{{r},v}^{s}(t) \cdot \chi_{r, v}^{{CPU}}(t)}{\zeta_{s}^{{CPU}}}, \quad s \in \mu
\end{equation}

Consequently, the total E2E delay for VNF-FG $r$ at time $t$ can be calculated as:
\begin{equation}
\begin{split}
D^{r}_{E2E}(t) =  \sum_{\bar{l} \in \bar{L_r},\; l \in L_{r}}  z_{r, \overline{l}}^{l}(t) \cdot D^{l}_{link}(t) \quad + \\ 
\sum_{{v} \in V_r,\; s \in \mu} y_{{r},v}^{s}(t) \cdot D^{s}_{proc}(t)
\end{split}
\end{equation}

\subsection {Energy Consumption Model}
The CPU is the primary source of server energy consumption, with its power usage assumed to scale linearly with utilization. Inspired by the energy consumption models presented in \cite{alharbi2019ant} and \cite{10509644},  the total energy consumption of server $s \in \mu$ that hosts VNF $v \in V_{r}$ from VNF-FG $r \in R$ at time step $t$ is given by:
\begin{equation}
E_{s}(t) = y_{{r},v}^{s}(t) \cdot \left(\varepsilon^{s}_{{base}}(t) + \varepsilon^{s}_{{cons}}(t) \cdot \Gamma^s_{CPU}(t)\right) + \varepsilon^{s}_{{trans}}(t), 
\end{equation}
here, $\varepsilon^{s}_{{base}}(t)$ denotes the baseline energy consumption of server $s$, while $\varepsilon^{s}_{{cons}}(t)$ represents the incremental energy consumption per unit of CPU utilization, defined as, $\varepsilon^{s}_{{cons}}(t)$ = $\varepsilon^{s}_{{max}}(t) - \varepsilon^{s}_{{base}}(t)$, where, $\varepsilon^{s}_{{max}}(t)$ denotes the maximum energy consumption when server $s$ operates at full CPU capacity. In addition, $\varepsilon^{s}_{{trans}}(t)$ represents the transition energy consumption of server $s$ at time step $t$. This transition energy consumption accounts for the power overhead associated with changes in the server's operational mode, such as transitioning from a sleep mode to an active mode due to VNF migration. If the operational mode of server $s$ remains unchanged between time steps $t-1$ and $t$, the transition energy consumption is assumed to be zero, that is, $\varepsilon^{s}_{{trans}}(t) = 0$.

\subsection {Problem Formulation}
The primary objective of this study is to maximize the reduction of both the E2E delay of VNF-FGs and the energy consumption in edge–core network through VNF migration. Consider, for example, a service request for 8K video streaming that is processed by a sequence of VNFs in a VNF-FG $r \in R$, such as a firewall (FW), traffic monitor (TM), and video optimization controller (VOC) at time step $t$. As illustrated in Figure 1, the E2E delay is reduced from 14 ms to 12 ms when the VOC is migrated from an edge cloud server $s_{e}$ to a core cloud server $s_{c}$. Furthermore, since edge cloud server $s_{e}$ no longer hosts any VNFs, it can be powered off, thereby contributing to a reduction in the overall energy consumption of the network.

The normalized reduction in E2E delay resulting from VNF migration is denoted as:
\begin{equation} 
\delta(t) = \frac{\Upsilon - \Upsilon'}\Upsilon
\end{equation}
where, $\Upsilon$ refers to the E2E delay before VNF migration, and $\Upsilon'$ denotes the E2E delay after migration.

Similarly, the normalized reduction in energy consumption due to VNF migration is given by:
\vspace{-0.5em}
\begin{equation} 
\eta(t) = \frac{\Lambda - \Lambda'}\Lambda
\end{equation}
where, $\Lambda$ represents the total energy consumption of the network before migration, and $\Lambda'$ denotes the energy consumption after migration.

Since the objective is to jointly maximize the reduction in both E2E delay and energy consumption, a unified variable is defined by combining the two metrics with corresponding weighting coefficients. This composite variable, derived from Eq. (14) and (15), is expressed as:
\begin{equation}
\lambda = (\tau_{1} \cdot \delta(t) + \tau_{2} \cdot \eta(t)),
\end{equation}
where, $\tau_{1}$ and $\tau_{2}$ are weighting coefficients in the range $[0, 1]$.  
\\
\textbf{Optimization 1.} \textit{For a set of $R$ VNF-FGs over time steps $t = 1$ to $T$, the VNF migration problem in the edge–core network is formulated as the maximization of the total average normalized reduction in E2E delay and energy consumption:}
\begin{equation}
\max_{\substack{}} \frac{1}{T} \sum_{t=1}^{T} \sum_{r \in {R}} \lambda , \quad \forall v \in V_{r}
\end{equation}
\text{subject to:}
\begin{equation}  
\textbf{C1}: \text{Eq.} (5) - \text{Eq.} (7), \ \ \ \ \
\end{equation}
\begin{equation}  
\textbf{C2}: D_{E2E}^{r}(t) \leq D_{max}^{r},  
\end{equation} 
here, $D_{max}^{r}$ denotes the maximum tolerable E2E delay for VNF-FG $r$.  

\section{DT-Empowered DRL for VNF Migration}
\subsection{Key Procedures of the Proposed VNF Migration  Scheme}
The proposed VNF migration scheme operates as follows, with detailed descriptions of the DRL module and the DT module presented in the following subsections.

At the beginning of each time step $t$, the system scans all servers and links to remove timed-out VNF-FGs, i.e., those for which $\Phi_{r}(t) = 0,\ \forall r \in R$. Subsequently, if there are incoming VNF-FG requests at time step $t$, the system accepts all of them and performs VNF deployment according to \cite{8905349}. Following this, the system employs the DT-empowered DRL framework to make decisions regarding VNF migration when necessary. Finally, the network configuration and the set of accepted requests are updated, marking the active VNF-FGs with $\Phi_{r}(t) = 1$.

\subsection{Description of the DRL Framework}
Since \textbf{Optimization 1} constitutes a joint optimization problem that aims to maximize the reduction of both E2E delay and average energy consumption, it poses significant challenges for conventional optimization techniques due to its complexity and inherent non-convexity. To address this, we reformulate \textbf{Optimization 1} as a Markov Decision Process (MDP) and adopt the A2C algorithm to solve it effectively. The formulated MDP can be represented as $\mathcal{M} = \{ \mathcal{S}, \mathcal{A}, \mathcal{R}, \mathcal{P} \}$, which includes the state space $\mathcal{S}$, the action space $\mathcal{A}$, the reward function $\mathcal{R}$, and the state transition probability $\mathcal{P}$. In practice, state transition probabilities are typically unknown, and A2C addresses this by learning through direct interaction with the environment.

In A2C, two separate deep neural networks, that is, actor and critic are implemented. The actor network outputs a stochastic policy $\pi(\mathcal{A}|\mathcal{S}; \theta^\pi)$, while the critic estimates the state value function $V(\mathcal{S}; \Theta^V)$. The training leverages the advantage function defined as, $A(\mathcal{S}, \mathcal{A}) = \mathcal{R} + \gamma V(\mathcal{S'}; \Theta^V) - V(\mathcal{S}; \Theta^V)$, which captures how much better an action $\mathcal{A}$ is compared to the expected value of the current state. This advantage term is used to guide policy updates, thereby reducing the variance of gradient estimates and accelerating convergence.

The actor loss is defined as:
\begin{equation}
L_{{actor}}(\Theta^\pi) = -\mathbb{E}_{\mathcal{S}, \mathcal{A}} \left[ \log \pi(\mathcal{A}|\mathcal{S}; \Theta^\pi) \cdot A(\mathcal{S}, \mathcal{A}) \right]
\end{equation}
that encourages the policy to increase the probability of advantageous actions. The critic loss is computed as the mean squared error between the predicted and target value:
\begin{equation}
L_{{critic}}(\Theta^V) = \mathbb{E}_{\mathcal{S}, \mathcal{R}, \mathcal{S}'} \left[ \left( \mathcal{R} + \gamma V(\mathcal{S}'; \Theta^V) - V(\mathcal{S}; \Theta^V) \right)^2 \right]
\end{equation}

This dual-network structure, combined with advantage-based updates, ensures stable and sample-efficient learning, making A2C well-suited for complex and high-dimensional optimization problems such as the one tackled in this study. The components of MDP, that is, state, action, and reward function are describe as follows: \\
\textit{\textbf{State:} At time step $t$, the system state is represented by a vector $\mathcal{S}(t)$, consisting of eight key components: $\{\xi(t), \psi(t), \rho(t), \iota(t), \varphi(t), \varrho(t), \varpi(t), \varsigma(t)\}$. Specifically, $\xi(t)$ indicates whether each server is in active or sleep mode, while $\psi(t)$ indicates whether each server is hosting at least one VNF, where $\psi(t) = y_{r,v}^{s}(t)$. The components $\rho(t)$, $\iota(t)$, $\varphi(t)$, and $\varrho(t)$ represent the sets of CPU utilizations, memory utilizations, energy consumptions, and processing delays, respectively, for all servers $s \in \mu$ at time step $t$. Additionally, $\varpi(t)$ captures the server-level deployment mapping of VNFs for each VNF-FG $r \in R$, specifying which servers are hosting VNFs from $r$ at time $t$. Finally, $\varsigma(t)$ denotes the set of resource demands, specifically, the memory and CPU requirements associated with each VNF within its corresponding VNF-FG $r$.}\\
\textit{\textbf{Action:} At time step $t$, the action taken by the system, denoted as $\mathcal{A}(t)$, represents the selection of servers from the set $\mu$ for migrating VNFs associated with a VNF-FG $r \in R$, where, $\mu = \mu_e \cup \mu_c$, with $\mu_e = \{s_{1}, s_{2}, \dots, s_{e}\}$ and $\mu_c = \{s_{e+1}, \dots, s_{c}\}$.}\\
\textit{\textbf{Reward Function:} At time step $t$, the reward function for migrating VNFs associated with a VNF-FG $r \in R$ is computed based on Eq. (16) as follows:}
\begin{equation}
\mathcal{R}(\mathcal{S}(t),\mathcal{A}(t)) = tanh (\lambda),
\end{equation}
here, ${tanh}(\cdot)$ denotes the hyperbolic tangent function which is applied to normalize the reward within a bounded range.

\begin{figure} [t]
\centerline{\includegraphics[width=1\columnwidth]{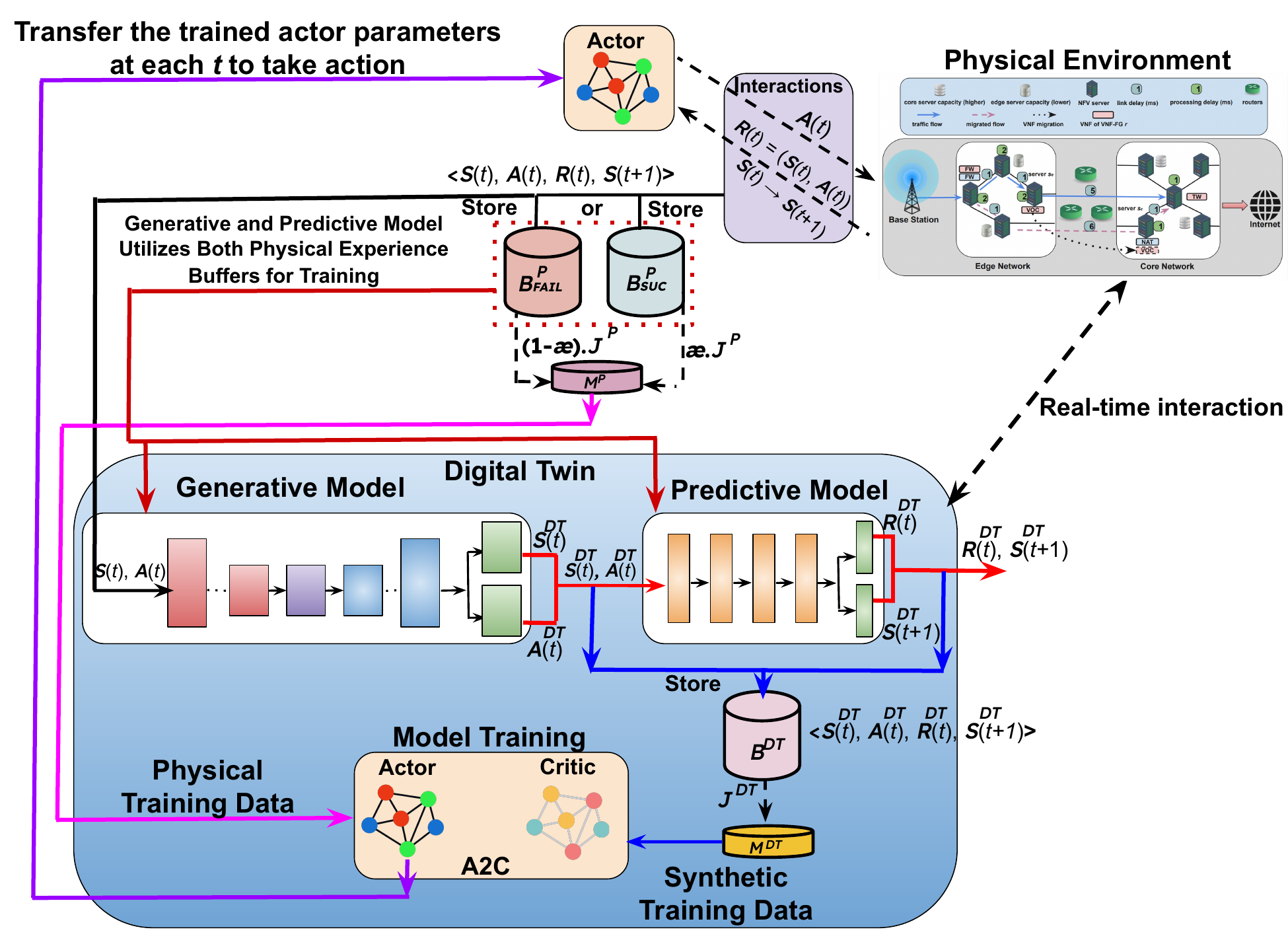}}
\caption{Operation of the DT-empowered DRL.}
\end{figure}

\subsection{Description of the DT Module}
In this study, a DT is established to enhance the performance and sustainability of the A2C model. Within the DT module, a multi-task VAE is implemented as the generative model to produce synthetic state–action pairs, and a multi-task LSTM network is utilized as the predictive model to estimate the corresponding next states and rewards. This combination enables the DT to emulate realistic environment dynamics and augment the learning process with high-quality synthetic data. Furthermore, by shifting the A2C training process from the physical environment to the DT, the framework significantly lowers the energy overhead and mitigates operational disruptions caused by frequent model updates.

At time step $t$, the generative model $\mathcal{V}$ takes as input the physical environment's state $\mathcal{S}(t)$ and action $\mathcal{A}(t)$, and generates the corresponding state and action in the DT, denoted as $\mathcal{S}^{{DT}}(t)$ and $\mathcal{A}^{{DT}}(t)$. This process can be formally expressed as:
\begin{equation} 
\mathcal{V}(\mathcal{S}(t), \mathcal{A}(t)) = \mathcal{S}^{{DT}}(t), \mathcal{A}^{{DT}}(t) 
\end{equation}

The loss function for the generative model is composed of a reconstruction term and a regularization term. It can be expressed as:
\begin{align}
\mathcal{L}(\theta, \phi) = 
& -\mathbb{E}_{z \sim q_\phi(z|x_{{in}})} \Big[ 
    \log p_\theta(\mathcal{S}^{DT} | z) 
    + \log p_\theta(\mathcal{A}^{DT} | z) \Big] \nonumber \\
& + \mathrm{KL}\left( q_\phi(z|x_{{in}}) \,\|\, p(z) \right),
\end{align}
where, $x_{{in}} = (\mathcal{S}(t), \mathcal{A}(t))$ is the input to the encoder, $q_\phi(z|x_{{in}})$ is the approximate posterior, $p_\theta(\cdot|z)$ are the decoder likelihoods for state and action outputs, and $p(z)$ is the prior distribution over latent variables. The Kullback–Leibler (KL) divergence term ensures that the learned latent distribution remains close to the prior.

Moreover, at time step $t$, the predictive model $\mathcal{T}$ utilizes the generated DT state $\mathcal{S}^{{DT}}(t)$ and action $\mathcal{A}^{{DT}}(t)$ as inputs to predict the next state $\mathcal{S}^{{DT}}(t+1)$ and the corresponding reward $\mathcal{R}^{{DT}}(t)$, and can be expressed as:
\begin{equation} 
\mathcal{T}(\mathcal{S}^{DT}(t), \mathcal{A}^{DT}(t)) = \mathcal{S}^{{DT}}(t+1), \mathcal{R}^{{DT}}(t) 
\end{equation}

The adopted loss function for the predictive model is the mean squared error (MSE), can be formulated as:  
\begin{equation}
\begin{aligned}
\mathcal{L}_{{MSE}} = 
&\; \kappa_{1} \cdot \left\| {\mathcal{S}}^{DT}(t+1) - \mathcal{S}(t+1) \right\|^2 \\
&+ \kappa_{2} \cdot \left\| {\mathcal{R}}^{DT}(t) - \mathcal{R}(\mathcal{S}(t), \mathcal{A}(t)) \right\|^2,
\end{aligned}
\end{equation}
where, $\kappa_{1}$ and $\kappa_{2}$ are weighting coefficients, ranging from $[0, 1]$.

A DT experience buffer, denoted as $\mathcal{B}^{DT}$ is established to store the generated data tuples, comprising $\langle \mathcal{S}^{{DT}}(t), \mathcal{A}^{{DT}}(t), \mathcal{R}(\mathcal{S}^{{DT}}(t), \mathcal{A}^{{DT}}(t)), \mathcal{S}^{{DT}}(t+1) \rangle$. 
This virtual experience buffer is utilized alongside the physical experience buffers to train the A2C, thereby improving sample efficiency and accelerating policy convergence.

\subsection{DT-Empowered DRL Framework Training}
To train the A2C, the DT experience buffer $\mathcal{B}^{DT}$ and two physical experience buffers are utilized, namely, the successful physical experience buffer $\mathcal{B}^{P}_{{SUC}}$ and the unsuccessful physical experience buffer $\mathcal{B}^{P}_{{FAIL}}$. Separating successful and unsuccessful experiences helps the system distinguish between constraint-satisfying and constraint-violating actions, enabling more informed policy updates than conventional random sampling. At each time step $t$, after executing an action, if the constraints \textbf{C1} and \textbf{C2} are not satisfied, the corresponding experience is stored in $\mathcal{B}^{P}_{{FAIL}}$; otherwise, it is stored in $\mathcal{B}^{P}_{{SUC}}$. During each training iteration, the system randomly samples $\varkappa \cdot J^{P}$ experiences from $\mathcal{B}^{P}_{{SUC}}$ and $(1-\varkappa) \cdot J^{P}$ experiences from $\mathcal{B}^{P}_{{FAIL}}$ to form the physical mini-batch $M^{P}$. The parameter $\varkappa \in [0, 1]$ controls the balance between successful and unsuccessful experiences and is critical for stable training.
In the case of $\mathcal{B}^{{DT}}$, a DT mini-batch $M^{{DT}}$ is formed by randomly sampling $J^{{DT}}$ experiences from the $\mathcal{B}^{{DT}}$.

During the training iterations, the combined mini-batches $M^{P}$ and $M^{{DT}}$ are used to update the A2C actor and critic networks. After training, the updated actor parameters are transferred back to the physical environment for action selection.
It is important to note that the generative and predictive models within the DT are trained using experiences drawn from $\mathcal{B}^{P}_{{FAIL}}$ and $\mathcal{B}^{P}_{{SUC}}$ at the beginning of each episode. 
The overall training and interaction process is illustrated in Figure 2.

\section{Performance Evaluation}

\subsection{Simulation Setup}
A physical edge–core network is generated based on the Waxman topology model with parameters $\alpha=0.5$ and $\beta=0.2$, following the configuration described in \cite{9060910}. The network consists of 100 servers, with 35 designated as edge cloud servers and 65 as core cloud servers, as specified in \cite{9440734}.
The edge cloud servers are configured with 40 units of CPU resources and 16 units of memory resources, whereas the core cloud servers are configured with 200 units of CPU resources and 64 units of memory resources \cite{9440734}. The CPU and memory utilization thresholds are both set to 80\%. Additionally, the available bandwidth for each link is set to 3.5 Gbps \cite{10729270}. The CPU and memory resource demands for each VNF within the VNF-FGs are randomly assigned, ranging from 1 to 20 units for CPU and from 1 to 4 units for memory, consistent with \cite{9440734}. 
Each VNF-FG is composed of four VNFs. In each episode, 300 VNF-FG requests arrive sequentially, following a Poisson process with an arrival rate of 0.2 requests per time step. 

In the A2C model, both the actor and critic networks are implemented as fully connected neural networks with four hidden layers comprising 1024, 512, 256, and 128 neurons, respectively, each using ReLU activation. The actor network generates a probability distribution over all candidate migration actions through a Softmax output layer. The critic network estimates the state-value using a single linear output neuron.
In the DT module, the predictive model comprises a single LSTM layer with 64 hidden units, followed by a fully connected layer with 64 ReLU-activated neurons. Its output is split into two task-specific heads: one predicts the next state, the other the immediate reward, both using linear activation. The generative model includes an encoder with two ReLU-activated hidden layers of 64 and 32 neurons, followed by two output layers estimating the mean and log-variance of the latent distribution. The decoder consists of a shared hidden layer with 64 ReLU-activated neurons and two task-specific Sigmoid heads for reconstructing the digital twin state and action, respectively. Table I summarizes the simulation parameters used for the A2C and associated neural networks.
\begin{table}[!h]
\centering
\caption{DRL and DT parameters for simulation}
\footnotesize 
\begin{tabularx}{\columnwidth}{Xcc}
\toprule
\textbf{Parameters} & \textbf{Values} \\
\midrule
Successful experience buffer size ($\mathcal{B}^{P}_{{SUC}}$) & 4000 \\ 
Unsuccessful experience buffer size ($\mathcal{B}^{P}_{{FAIL}}$) & 2000 \\
Virtual experience buffer size ($\mathcal{B}^{DT}$) & 6000 \\
Balance parameter ($\varkappa$) & 0.30 \\
Learning rate of actor \& critic & 0.1 \\
Discount factor ($\gamma$) & 0.95 \\
Physical mini-batch size  ($M^{{P}}$) & 32 \\
DT mini-batch size  ($M^{{DT}}$) & 32 \\
Optimizer & $Adam$\\ 
\bottomrule
\end{tabularx}
\end{table}

Furthermore, the Tensorflow-Keras platform is utilized to
construct and train DNN models. Simulations are executed on a 2021 MacBook Pro, equipped with an Apple M1 Pro chip, featuring a 64-bit operating system, an 8-core CPU, a 14-core GPU, and 16GB of unified RAM. 

For benchmarking performance comparisons, two VNF migration schemes are evaluated: (1) a non–DT-based A2C VNF migration and (2) a random VNF migration. 

\subsection{Numerical Results and Comparisons}

\begin{figure} 
    \centering
  \subfloat[\label{1a}]{%
       \includegraphics[width=0.495\linewidth]{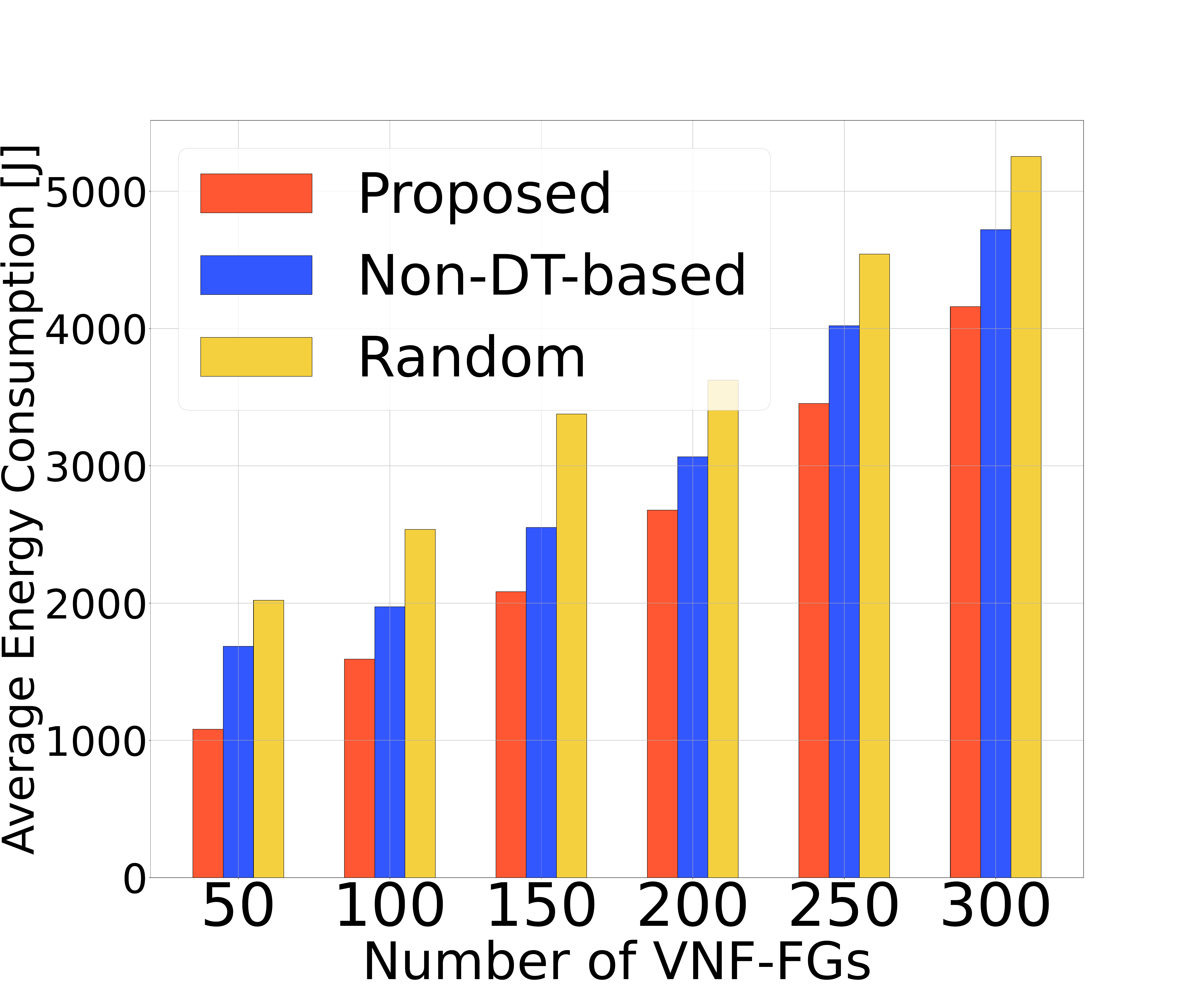}}
    \hfill  
    \subfloat[\label{1b}]{%
        \includegraphics[width=0.49\linewidth]{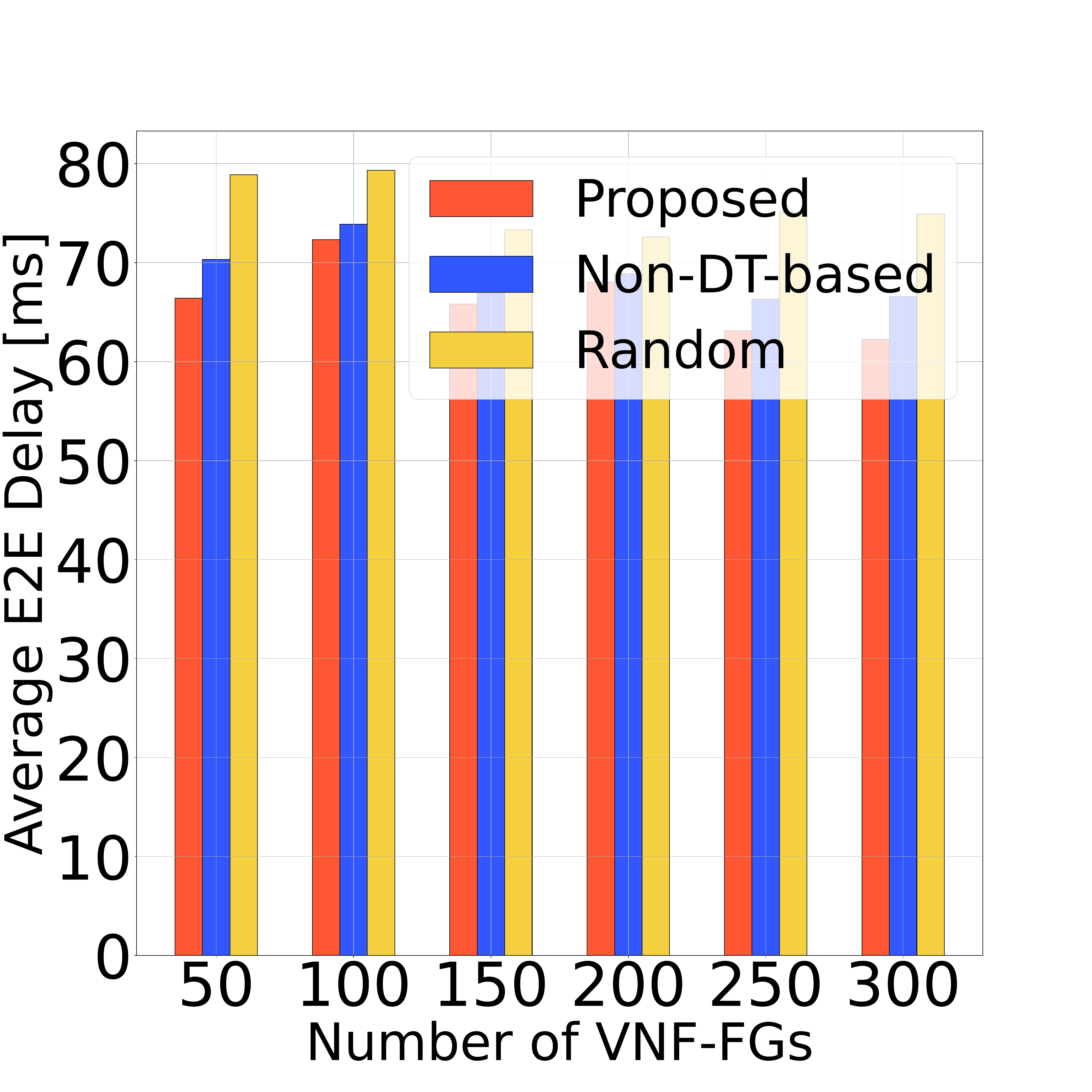}}
        \vspace{-10pt}
        \hfill
  \subfloat[\label{1c}]{%
        \includegraphics[width=0.6\linewidth]{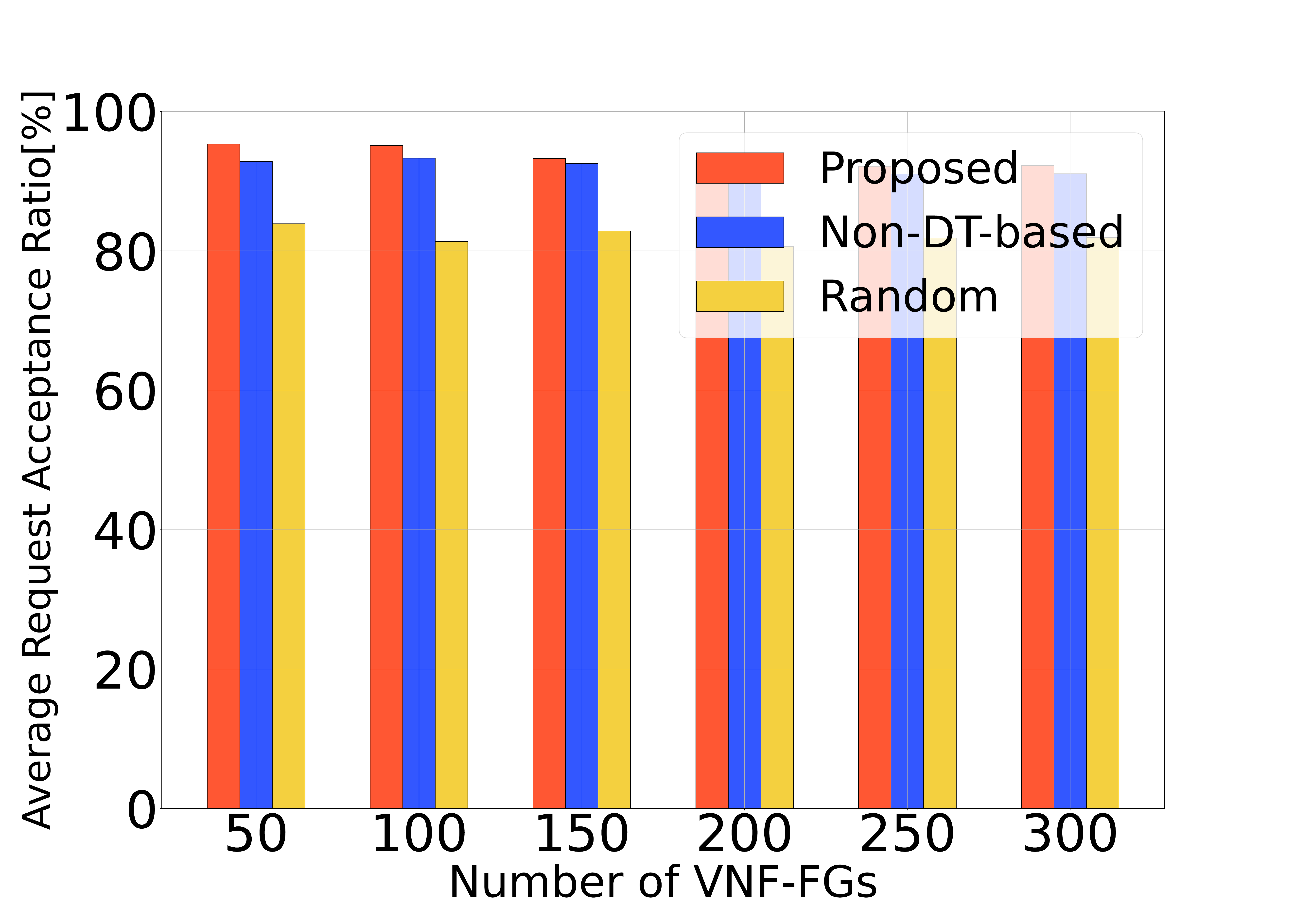}}
  \caption{(a) Average energy consumption vs. number of VNF-FGs, (b) Average E2E delay vs. number of VNF-FGs, (c) Average request acceptance ratio vs. number of VNF-FGs.}
  \label{fig3}
\end{figure}

Figure 3(a) illustrates that the proposed framework outperforms baseline approaches, such as the non–DT-based A2C VNF migration and random VNF migration scheme, in terms of average energy consumption across varying numbers of VNF-FGs. Specifically, it achieves reductions of 15\% and 27\%, respectively. This gain results from integrating A2C with the DT module, which provides synthetic constraint-aware transitions that improve the A2C’s ability to identify energy-efficient migration decisions. By exposing the policy to a broader range of system states, the DT helps avoid unnecessary server activations and accelerates convergence toward energy-minimizing behavior.

Figure 3(b) presents the average E2E delay of the proposed framework compared to baseline approaches, such as the non–DT-based A2C VNF migration and random VNF migration scheme. The proposed method achieves the lowest average E2E delay, with reductions of 7\% and 18\%, respectively. This improvement stems from the DT-generated constraint-aware synthetic transitions, which offer richer exposure to delay-critical states. Consequently, the policy avoids delay inducing actions such as migrating VNFs to overloaded servers and converges toward more delay-efficient migration strategies.

Figure 3(c) illustrates that the proposed framework achieves a higher average request acceptance ratio compared to both the non–DT-based A2C VNF migration and the random VNF migration scheme. This improvement is attributed to the DT–assisted learning process which enables the policy to anticipate resource contention and avoid infeasible migration decisions by leveraging synthetic, constraint-aware experiences, thereby maintaining higher service admission rates under dynamic network conditions.

\section{Conclusions }
\label{sec:Conclusions}
In this study, a novel DT-assisted DRL framework is proposed to enhance the efficiency of VNF migration in edge–core networks. By integrating the A2C model with a DT that captures network dynamics through a multi-task LSTM network and a multi-task VAE, the proposed solution effectively achieves the objective of minimizing both average end-to-end delay and energy consumption. The incorporation of synthetic experience buffers alongside physical experience buffers further improves sample efficiency, enabling robust training even under highly dynamic workload conditions. Extensive evaluation demonstrates that the proposed framework consistently outperforms traditional non–DT-based VNF migration and random VNF migration schemes across key performance metrics, such as average energy consumption, average E2E delay, and acceptance ratio. These findings validate the effectiveness of leveraging DT and DRL to enable scalable, adaptive, and sustainable orchestration of VNFs.


\bibliographystyle{IEEEtran}
\bibliography{IEEEabrv,bib}
\end{document}